\documentclass[11pt,a4paper]{article}

\usepackage[utf8]{inputenc}
\usepackage[T1]{fontenc}
\usepackage{times}

\usepackage[margin=1in]{geometry}
\usepackage{amsmath,amssymb}
\usepackage{booktabs}
\usepackage{graphicx}
\usepackage{subcaption}
\usepackage{xcolor}
\usepackage{listings}
\usepackage{enumitem}
\usepackage{hyperref}
\usepackage{url}
\usepackage{natbib}
\usepackage{authblk}
\usepackage{float}
\usepackage{multirow}
\usepackage{tabularx}

\hypersetup{
  colorlinks=true,
  linkcolor=blue!70!black,
  citecolor=blue!70!black,
  urlcolor=blue!70!black,
}

\lstdefinestyle{protocol}{
  basicstyle=\ttfamily\small,
  breaklines=true,
  frame=single,
  backgroundcolor=\color{gray!5},
  columns=fullflexible,
  keepspaces=true,
  numbers=none,
  xleftmargin=3pt,
  xrightmargin=3pt,
}

\newcommand{\aip}{\textsc{AIP}}
\newcommand{\ibct}{\textsc{IBCT}}
\newcommand{\ldp}{\textsc{LDP}}
\newcommand{\dci}{\textsc{DCI}}
\newcommand{\atoa}{\textsc{A2A}}
\newcommand{\mcp}{\textsc{MCP}}
\newcommand{\jamjet}{\textsc{JamJet}}

\begin{document}

\title{AIP: Agent Identity Protocol for Verifiable Delegation Across MCP and A2A}

\author[1]{Sunil Prakash}
\affil[1]{Indian School of Business, India \\ \textit{sunil\_prakash\_pgpmax2026@isb.edu}}
\date{}

\maketitle

\begin{abstract}
AI agents increasingly call tools via the Model Context Protocol (\mcp{}) and delegate to other agents via Agent-to-Agent (\atoa{}), yet neither protocol verifies agent identity. A scan of approximately 2,000 \mcp{} servers found all lacked authentication. In our survey of eleven categories of prior work, we did not identify a prior \textit{implemented} protocol that jointly combines public-key verifiable delegation, holder-side attenuation, expressive chained policy, transport bindings across MCP/A2A/HTTP, and provenance-oriented completion records. We introduce Invocation-Bound Capability Tokens (\ibct{}s), a primitive that fuses identity, attenuated authorization, and provenance binding into a single append-only token chain. \ibct{}s operate in two wire formats: compact mode (a signed JWT for single-hop cases) and chained mode (a Biscuit token with Datalog policies for multi-hop delegation). We provide reference implementations in Python and Rust with full cross-language interoperability. Compact mode verification takes 0.049\,ms (Rust) and 0.189\,ms (Python). Chained tokens scale linearly at 340--380 bytes per delegation block, with sub-millisecond verification even at depth~5. In a real MCP deployment over HTTP, \aip{} adds 0.22\,ms overhead over no-auth baselines. In a real multi-agent deployment with Gemini 2.5 Flash, \aip{} adds 2.35\,ms of overhead (0.086\% of total latency). Adversarial evaluation across 600 attack attempts shows 100\% rejection, with two attack categories (delegation depth violation and audit evasion) uniquely caught by \aip{}'s chained delegation model.
\end{abstract}

\section{Introduction}
\label{sec:intro}

AI agents are acquiring the ability to act: calling tools, spending money, and delegating work to other agents. The two dominant protocols enabling this, the Model Context Protocol (\mcp{})~\citep{anthropic2024mcp} for tool invocation and Agent-to-Agent (\atoa{})~\citep{google2025a2a} for inter-agent collaboration, define how agents communicate but not who they are. \mcp{} adopted OAuth 2.1 as an optional authorization layer in 2026, yet a Knostic security scan~\citep{knostic2025scan} of approximately 2,000 \mcp{} servers found that every single one lacked authentication. \atoa{} agent cards contain self-declared identities with no attestation binding. When Agent~A delegates a task to Agent~B, no mechanism verifies A's authority, constrains B's scope, or records the delegation for audit.

Recent work addresses fragments of this gap. Google DeepMind's ``Intelligent AI Delegation'' paper~\citep{deepmind2026delegation} proposed Delegation Capability Tokens built on macaroons, validating the attenuation-first design philosophy but shipping no protocol or implementation. Four IETF Internet-Drafts published in early 2026 (AIMS~\citep{ietf2026aims}, WIMSE~\citep{ietf2026wimse}, Agentic JWT~\citep{ietf2026agenticjwt}, SCIM for agents~\citep{ietf2026scim}) target individual aspects. Mastercard's Verifiable Intent proposal~\citep{mastercard2025intent} targets agent commerce. Several partial solutions have emerged, but in our survey we did not find a single \textit{implemented} protocol that combines offline attenuable delegation, expressive chained policy, provenance-aware completion records, and transport bindings across MCP, A2A, and HTTP.

Existing identity and authorization approaches each fail at least one requirement of agent delegation. OAuth 2.0/2.1~\citep{ietf2024oauth21} requires centralized authorization servers, produces opaque tokens with no delegation chain, and provides no holder-side scope attenuation. W3C DIDs~\citep{w3c2026did} introduce blockchain dependencies and circular trust bootstrapping. Macaroons~\citep{birgisson2014macaroons} use HMAC shared secrets that make every verifier a potential forger. UCANs~\citep{ucan2024spec} inherit DID complexity and suffer quadratic token bloat from nested JWTs. SPIFFE/SPIRE~\citep{spiffe2024} demands dedicated infrastructure incompatible with ephemeral agent creation. Biscuit~\citep{biscuit2024spec} gets the token format right but defines no identity resolution, no protocol bindings, and no provenance model. Agent identity requires seven properties (Section~\ref{sec:gap}); no single solution in our survey addresses more than four.

This paper makes two contributions. First, we introduce Invocation-Bound Capability Tokens (\ibct{}s), a primitive that unifies identity, attenuated authorization, and provenance binding in a single evolvable token chain. \ibct{}s operate in two wire formats: compact mode (a signed JWT for single-hop \mcp{} calls) and chained mode (a Biscuit token with Datalog policies for multi-hop delegation chains with completion blocks). Second, we evaluate \aip{} with reference implementations in Python and Rust, demonstrating sub-millisecond verification for compact tokens (0.049\,ms Rust, 0.189\,ms Python), linear scaling of chained tokens at 340--380 bytes per delegation block, negligible overhead (0.086\% of total latency) in a real multi-agent deployment with Gemini 2.5 Flash inference, and 100\% attack rejection across 600 adversarial tests spanning six attack categories.

\aip{} complements our prior work on \ldp{}~\citep{prakash2025ldp} (provenance) and \dci{}~\citep{prakash2025dci} (reasoning), addressing the identity layer of the agent trust stack. The three protocols are independently deployable; integration points are described in Section~\ref{sec:impl}.

The remainder of this paper is organized as follows. Section~\ref{sec:related} surveys related work and identifies the seven-property gap. Section~\ref{sec:protocol} defines the \aip{} protocol. Section~\ref{sec:impl} describes the implementation. Section~\ref{sec:eval} evaluates performance and security. Section~\ref{sec:threat} presents the threat model. Section~\ref{sec:limitations} discusses limitations. Section~\ref{sec:conclusion} concludes.

\section{Related Work and Gap Analysis}
\label{sec:related}

Agent identity sits at the intersection of decentralized identity, capability-based authorization, and service mesh security. This section surveys eleven categories of prior work, identifies the specific property each fails to provide, and synthesizes the gap that \aip{} addresses. Table~\ref{tab:comparison} summarizes the analysis across seven dimensions.

\subsection{W3C Decentralized Identifiers}
\label{sec:did}

W3C Decentralized Identifiers (DIDs)~\citep{w3c2026did} provide self-sovereign, blockchain-anchored identity. DID v1.1 reached Candidate Recommendation in March 2026, yet adoption remains limited: Block abandoned its ``Web5'' DID initiative in late 2024 after failing to overcome wallet UX friction. For AI agents, wallet UX is irrelevant (agents are software), but the structural problems persist. Most DID methods depend on blockchain infrastructure for resolution, introducing latency and availability concerns incompatible with high-throughput agent delegation. Trust bootstrapping remains circular: a DID is self-certifying, but verifiers need an external reason to trust it, reintroducing centralized trust anchors. For human identity where self-sovereignty and censorship resistance matter, DIDs provide properties that \aip{} does not target.

\subsection{OAuth 2.0/2.1}
\label{sec:oauth}

OAuth 2.1~\citep{ietf2024oauth21} is the most widely deployed authorization framework. \mcp{} adopted OAuth 2.1 with PKCE in 2026~\citep{anthropic2024mcp}, and \atoa{} supports RFC 8693 token exchange~\citep{google2025a2a}. For single-domain human-to-service authentication, OAuth remains the better choice due to its mature ecosystem, universal library support, and well-understood security properties. Three structural limitations prevent it from serving as the agent identity layer. First, every trust domain requires its own authorization server, with no mechanism for cross-domain token verification without pre-established federation. Second, OAuth tokens are opaque to intermediaries: delegation via token exchange produces a new token with no record of the original authorization chain. Third, OAuth provides no holder-side scope attenuation; only the authorization server can narrow scopes at issuance time.

\subsection{Macaroons}
\label{sec:macaroons}

Macaroons~\citep{birgisson2014macaroons} introduced holder attenuation as a first-class primitive. Google DeepMind's ``Intelligent AI Delegation'' paper~\citep{deepmind2026delegation} builds directly on macaroons, proposing Delegation Capability Tokens (DCTs) for agent-to-agent delegation. For simple attenuation scenarios within a single trust domain where all verifiers are trusted, macaroons provide a lightweight and well-studied primitive with minimal token overhead. Three limitations prevent broader deployment. First, HMAC-based verification requires the verifier to possess the root secret, making every verifier a potential forger. Second, macaroon caveats are simple key-value predicates, insufficient for expressing conditional policies (temporal bounds, budget limits, tool-parameter constraints). Third, DeepMind's DCT proposal remains a position paper with no shipped protocol or reference implementation.

\subsection{UCAN}
\label{sec:ucan}

User Controlled Authorization Networks (UCANs)~\citep{ucan2024spec} implement decentralized delegation via nested JWTs with DID-based identity. For web3-native applications where DID infrastructure is already deployed, UCANs provide a well-specified delegation model with an active open-source ecosystem. Four limitations apply to the agent delegation use case. First, DID dependency inherits blockchain infrastructure requirements (Section~\ref{sec:did}). Second, nested JWT encoding creates token bloat that grows quadratically with delegation depth. Third, the UCAN ecosystem is JS-centric and tightly coupled to web3 storage platforms. Fourth, UCAN capabilities are flat URIs (\texttt{msg/send}, \texttt{file/read}), lacking expressiveness for conditional policies, budget constraints, or temporal restrictions.

\subsection{Biscuit}
\label{sec:biscuit}

Biscuit~\citep{biscuit2024spec} combines Ed25519 public-key signatures, a Datalog policy language, and an append-only block structure. Among existing solutions, Biscuit gets the most right: public-key verification eliminates shared secrets, Datalog provides expressive policy evaluation, and append-only blocks naturally model delegation chains. For applications that need only a token format without identity resolution or protocol bindings, raw Biscuit provides a simpler integration path with well-tested libraries in Rust, Go, Java, and Python. However, Biscuit is a token format, not an identity protocol. It defines no identity resolution, no protocol bindings for \mcp{}/\atoa{}/HTTP, and no provenance concept. Two independent deployments using raw Biscuit tokens have no interoperability path. Section~\ref{sec:biscuit-delta} details what \aip{} adds on top of Biscuit's cryptographic foundation.

\subsection{SPIFFE/SPIRE}
\label{sec:spiffe}

SPIFFE~\citep{spiffe2024} provides workload identity in cloud-native environments via X.509 or JWT SVIDs, with production deployments at Uber, Stripe, and Netflix. For static infrastructure workloads within Kubernetes clusters, SPIFFE provides stronger guarantees than \aip{} through hardware attestation, automatic certificate rotation, and deep integration with service mesh platforms. Three factors limit its applicability to agent identity. First, SPIRE requires dedicated infrastructure (attestation nodes, registration servers, certificate authorities). Second, X.509 certificate issuance latency is incompatible with ephemeral agent creation. Third, SPIFFE addresses workload identity within infrastructure boundaries but provides no cross-protocol identity flow: a SPIFFE SVID for an \mcp{} tool call has no mapping to the \atoa{} agent card.

\subsection{IETF Drafts}
\label{sec:ietf}

Four IETF Internet-Drafts published in early 2026 address agent identity. AIMS~\citep{ietf2026aims} proposes a conceptual model composing WIMSE, SPIFFE, and OAuth, but its authorization section reads ``TODO Security.'' WIMSE~\citep{ietf2026wimse} introduces a Dual-Identity Credential binding agent identity to owner identity, solving two-party interactions but not multi-hop delegation. Agentic JWT~\citep{ietf2026agenticjwt} extends JWTs with agent-specific claims, but JWTs are immutable after signing, so a delegatee cannot attenuate authority without minting a new token that breaks the cryptographic chain. SCIM for agents~\citep{ietf2026scim} addresses provisioning lifecycle rather than runtime authorization. For organizations already invested in IETF-standard infrastructure, these drafts offer the advantage of alignment with existing enterprise identity stacks. Collectively, however, no single draft provides holder-attenuable delegation, cross-protocol bindings, and provenance tracking in a unified protocol.

\subsection{Agent Authorization Profile (AAP) for OAuth 2.0}
\label{sec:aap}

The Agent Authorization Profile (AAP)~\citep{ietf2026aap} extends OAuth 2.0 and JWT with structured claims for agent identity, task binding, capabilities, delegation chains, and oversight. Rather than introducing a new protocol, AAP profiles existing OAuth infrastructure, which lowers adoption barriers for organizations already invested in OAuth. AAP gets several things right: it defines agent-specific JWT claims, supports delegation chain representation, and reuses well-understood OAuth flows. For organizations with existing OAuth infrastructure that need agent-aware claims without adopting a new protocol, AAP offers the lowest migration cost. Three limitations apply to the multi-hop agent delegation use case. First, JWT-based tokens are immutable after signing, so a holder cannot attenuate authority offline without minting a new token that breaks the cryptographic chain. Second, AAP requires an OAuth authorization server for every trust domain, precluding lightweight cross-domain delegation. Third, AAP defines no cross-protocol bindings for MCP or A2A and no provenance or completion concept.

\subsection{Transaction Tokens for Agents}
\label{sec:txntokens}

The IETF draft on Transaction Tokens for Agents~\citep{ietf2026txntokens} extends OAuth Transaction Tokens with \texttt{actor} and \texttt{principal} fields for agent context propagation across service boundaries. This approach solves an important problem: preserving the identity of the original principal as requests traverse multiple services within a trust domain. For within-domain service chains where the primary concern is principal propagation, Transaction Tokens offer a well-scoped solution with minimal protocol overhead. Two limitations prevent them from serving as a general agent delegation protocol. First, Transaction Tokens are designed for within-trust-domain chaining and do not address cross-domain delegation where no shared authorization server exists. Second, the specification defines no scope attenuation semantics (a downstream service receives the same authority as the upstream caller) and no completion or provenance binding.

\subsection{Agent Identity and Discovery (AID)}
\label{sec:aid}

The Agent Identity and Discovery (AID) protocol~\citep{ietf2026aid} proposes DNS-first discovery using TXT records at \texttt{\_agent.<domain>}, with optional public-key attestation. AID addresses a complementary layer: discovery (``where is the agent?'') rather than authorization (``what can it do?''). For the pure discovery problem, AID is the better fit, providing a lightweight DNS-native mechanism without the overhead of a full authorization protocol. Both AID and AIP use DNS as a trust anchor but at different protocol layers. AID could serve as the discovery mechanism that directs verifiers to an AIP identity document, making the two protocols composable rather than competing.

\subsection{MCP Authorization (OAuth 2.1)}
\label{sec:mcp-auth}

The MCP specification now includes OAuth 2.1 with PKCE for protected MCP servers~\citep{anthropic2024mcp}, meaning MCP is no longer without an authentication story. For single-hop client-to-server authentication, MCP's OAuth integration is the simpler and more mature option with broader library support. \aip{}'s positioning must be precise: MCP has client-to-server authentication via OAuth, but OAuth tokens carry no delegation chain, provide no holder-side scope attenuation, and include no provenance binding. In a multi-hop scenario where an orchestrator delegates to a specialist that calls an MCP tool, OAuth authenticates the final hop but provides no record of the authorization chain that led to it. AIP and MCP OAuth are complementary: OAuth authenticates the transport connection; AIP authenticates the delegation chain.

\subsection{Gap Summary}
\label{sec:gap}

The preceding review reveals that agent identity requires seven properties. In our survey, we did not identify a single prior solution that addresses all seven simultaneously:

\begin{enumerate}[leftmargin=*,nosep]
  \item \textbf{Public-key verification} without shared secrets (macaroons, OAuth fail).
  \item \textbf{Holder-side scope attenuation} without contacting the issuer (OAuth, JWTs fail).
  \item \textbf{Expressive policy language} beyond flat scopes (macaroons, UCAN fail).
  \item \textbf{Cross-protocol identity flow} across \mcp{}, \atoa{}, and HTTP (all surveyed solutions address at most one transport).
  \item \textbf{Provenance binding} of outcomes to authorization chains (we did not find a surveyed solution that provides this).
  \item \textbf{No blockchain or heavy infrastructure dependency} (DIDs, SPIFFE fail).
  \item \textbf{Agent lifecycle awareness} for ephemeral agents with short-lived grants (OAuth, SPIFFE fail).
\end{enumerate}

Table~\ref{tab:comparison} maps each surveyed approach against these seven properties. No single solution in our survey addresses more than four. \aip{} is designed to satisfy all seven.

\begin{table*}[t]
\centering
\caption{Seven-dimension comparison of identity and authorization approaches for AI agents. \checkmark{} = fully supported, $\sim$ = partially supported, $\times$ = not supported.}
\label{tab:comparison}
\footnotesize
\setlength{\tabcolsep}{4pt}
\begin{tabular}{@{}lccccccc@{}}
\toprule
\textbf{Approach} & \textbf{Public-Key} & \textbf{Holder} & \textbf{Expressive} & \textbf{Cross-} & \textbf{Provenance} & \textbf{No Blockchain/} & \textbf{Lifecycle} \\
 & \textbf{Verif.} & \textbf{Attenuation} & \textbf{Policy} & \textbf{Protocol} & \textbf{Binding} & \textbf{Heavy Infra} & \textbf{Awareness} \\
\midrule
W3C DIDs        & \checkmark & $\times$ & $\times$ & $\times$ & $\times$ & $\times$ & $\times$ \\
OAuth 2.0/2.1   & $\sim$     & $\times$ & $\times$ & $\times$ & $\times$ & \checkmark & $\times$ \\
Macaroons       & $\times$   & \checkmark & $\times$ & $\times$ & $\times$ & \checkmark & $\times$ \\
UCAN            & \checkmark & \checkmark & $\times$ & $\times$ & $\times$ & $\times$ & $\times$ \\
Biscuit         & \checkmark & \checkmark & \checkmark & $\times$ & $\times$ & \checkmark & $\times$ \\
SPIFFE/SPIRE    & \checkmark & $\times$ & $\times$ & $\times$ & $\times$ & $\times$ & $\sim$ \\
IETF Drafts     & $\sim$     & $\times$ & $\times$ & $\times$ & $\times$ & $\sim$ & $\sim$ \\
AAP (OAuth)     & $\sim$     & $\times$ & $\sim$   & $\times$ & $\times$ & \checkmark & $\sim$ \\
Txn Tokens      & $\sim$     & $\times$ & $\times$ & $\times$ & $\times$ & \checkmark & $\times$ \\
AID             & $\sim$     & $\times$ & $\times$ & $\times$ & $\times$ & \checkmark & $\times$ \\
\midrule
\textbf{\aip{}} & \checkmark & \checkmark & \checkmark & \checkmark & \checkmark & \checkmark & \checkmark \\
\bottomrule
\end{tabular}
\vspace{0.5em}

\noindent\textit{Scoring criteria.} A checkmark (\checkmark) indicates that the approach provides this property as a core, implemented feature. A tilde ($\sim$) indicates that the approach partially addresses this property or could be extended to cover it. A cross ($\times$) indicates that the approach does not provide this property in its current specification or implementation.
\end{table*}

\subsection{Relationship to Biscuit}
\label{sec:biscuit-delta}

\aip{} uses Biscuit~\citep{biscuit2024spec} as the cryptographic primitive for chained mode, inheriting its append-only block structure, Ed25519 signature chaining, and Datalog policy evaluation. What \aip{} adds is everything required to turn a token format into an identity protocol: (1)~identity resolution (\texttt{aip:web} and \texttt{aip:key}), (2)~protocol bindings for \mcp{}, \atoa{}, and HTTP, (3)~completion blocks for provenance binding, (4)~policy profiles (Simple, Standard, Advanced), and (5)~a compact JWT mode for single-hop cases. \aip{} inherits Biscuit's formal security properties while addressing the protocol-level gaps that prevent raw Biscuit tokens from interoperating across agent ecosystems.

\section{The Agent Identity Protocol}
\label{sec:protocol}

\subsection{Identity Scheme}
\label{sec:identity}

\aip{} defines two identity resolution methods that share a common token format and verification path.

\paragraph{DNS-based identities} use the scheme \texttt{aip:web:<domain>/<path>} and resolve via HTTPS to a well-known endpoint. The identifier \texttt{aip:web:jamjet.dev/agents/research-analyst} resolves to \texttt{https://jamjet.dev/.well-known/aip/agents/research-analyst.json}. DNS-based identities are appropriate for long-lived agents operated by organizations that control a domain name.

\paragraph{Self-certifying identities} use the scheme \texttt{aip:key:ed25519:<multibase>}, where the identifier is the public key itself. No resolution step is required: the verifier extracts the key directly from the identifier. Self-certifying identities are appropriate for ephemeral sub-agents spawned by an orchestrator for a single task (Section~\ref{sec:delegation}).

Both resolution methods produce an \aip{} identity document: a JSON object containing the agent's public keys, delegation parameters, supported protocol bindings, and an expiration timestamp. Figure~\ref{fig:identity-doc} shows an abbreviated example.

\begin{figure}[t]
\begin{lstlisting}[style=protocol,caption={Abbreviated \aip{} identity document. The \texttt{document\_signature} covers the RFC~8785 canonical form of all other fields.},label={fig:identity-doc}]
{ "aip": "1.0",
  "id": "aip:web:jamjet.dev/agents/research",
  "public_keys": [{"id": "key-1",
    "type": "Ed25519",
    "public_key_multibase": "z6Mkf5rG...",
    "valid_from": "2026-03-01T00:00:00Z",
    "valid_until": "2026-06-01T00:00:00Z"}],
  "delegation": {"max_depth": 3,
    "allow_ephemeral_grants": true},
  "protocols": {"mcp": {"header": "X-AIP-Token"},
    "a2a": {"agent_card_field": "aip_identity"}},
  "document_signature": "<Ed25519 over RFC 8785>",
  "expires": "2026-06-22T00:00:00Z" }
\end{lstlisting}
\end{figure}

Identity documents MUST include a \texttt{document\_signature} field: an Ed25519 signature over the RFC~8785~\citep{rfc8785} canonical form (excluding the signature field itself). This self-signature provides content integrity independent of transport security, detecting tampering even if the hosting domain is compromised. Implementations reject documents with unsupported major versions and ignore unknown fields for forward compatibility.

\subsection{Invocation-Bound Capability Tokens}
\label{sec:ibct}

The core contribution of \aip{} is the \ibct{}: a single token that fuses identity, attenuated authorization, and provenance into an append-only chain. An \ibct{} evolves through three block types as delegation proceeds.

\paragraph{Block 0 (Authority).} The root block is signed by the human or system that initiates the delegation chain. It declares the root identity, initial capability scopes, a budget ceiling, the maximum delegation depth, and an expiry timestamp. This block establishes the ceiling of authority for the entire chain.

\paragraph{Block N (Delegation).} Each intermediary agent appends a delegation block, signed with its own key. A delegation block names the delegator and delegatee, attenuates scope (narrowing capabilities from those granted by the parent block), and includes a mandatory \texttt{context} field describing the purpose of the delegation. Scope attenuation is cryptographically enforced: a delegation block that attempts to widen any capability beyond its parent block fails verification.

\paragraph{Block N+1 (Completion).} When work finishes, the executing agent may append a completion block recording the result hash, verification status, resource consumption, and cost. Completion blocks are optional; their trust semantics are detailed in Section~\ref{sec:completion}.

A completed \ibct{} answers five questions from a single artifact: Who authorized this action? Through which agents did the delegation flow? What constraints applied at each hop? What was the outcome? Was the outcome independently verified?

\subsubsection{Wire Formats}

\aip{} defines two wire formats that serve different deployment scenarios.

\paragraph{Compact mode} uses a standard JWT (\texttt{alg: EdDSA}, \texttt{typ: aip+jwt}) with \aip{}-specific claims: \texttt{iss}, \texttt{sub}, \texttt{scope}, \texttt{budget\_usd}, \texttt{max\_depth}, and standard timestamps. Compact mode covers single-hop cases. Any JWT library supporting EdDSA can create and verify compact tokens.

\paragraph{Chained mode} uses the Biscuit token format~\citep{biscuit2024spec} with Ed25519-signed append-only blocks and Datalog policy evaluation. Chained mode is required for multi-hop delegation, cross-organization workflows, and audit trail scenarios.

\paragraph{Mode detection and upgrade.} Receivers detect the format from the token header (\texttt{typ: aip+jwt} vs. Biscuit magic bytes). Upgrading from compact to chained requires re-issuance, mapping JWT claims to Biscuit authority facts per Table~\ref{tab:claim-map}.

\begin{table}[t]
\centering
\caption{Compact-to-chained claim mapping. Each JWT claim maps to a Biscuit authority fact during token upgrade.}
\label{tab:claim-map}
\footnotesize
\begin{tabular}{@{}ll@{}}
\toprule
\textbf{JWT Claim} & \textbf{Biscuit Fact} \\
\midrule
\texttt{iss} & \texttt{identity(\$iss)} \\
\texttt{sub} & \texttt{delegate(\$sub)} \\
\texttt{scope[i]} & \texttt{right(\$scope\_item)} \\
\texttt{budget\_usd} & \texttt{budget(\$budget\_usd)} \\
\texttt{max\_depth} & \texttt{max\_depth(\$max\_depth)} \\
\texttt{exp} & \texttt{expires(\$exp)} \\
\bottomrule
\end{tabular}
\end{table}

Figure~\ref{fig:delegation-chain} illustrates a three-hop chained token with scope narrowing at each delegation step and a completion block at the end.

\begin{figure}[t]
\begin{lstlisting}[style=protocol,caption={Three-hop delegation chain with scope narrowing and completion block.},label={fig:delegation-chain}]
Block 0 (Authority) -- signed by root
  identity: aip:web:acme.dev/orchestrator
  scope: [tool:*, delegate:*, budget:5.00], max_depth: 3
Block 1 (Delegation) -- signed by orchestrator
  delegate: aip:web:acme.dev/research-analyst
  scope: [tool:search, tool:browse, budget:0.50]
Block 2 (Delegation) -- signed by research-analyst
  delegate: aip:key:ed25519:z6Mkf...
  scope: [tool:search, budget:0.10]
Block 3 (Completion) -- signed by ephemeral agent
  status: completed, result_hash: sha256:e3b0c4...
  cost_usd: 0.03, tokens_used: 1200
\end{lstlisting}
\end{figure}

\subsection{Completion Blocks and Trust Model}
\label{sec:completion}

Completion blocks are signed by the executing agent, making them \textbf{self-reported claims}: tamper-evident and attribution-bound, but not independently verified by default. \aip{} defines three trust escalation levels: (1)~\textbf{self-reported} (default), where the executing agent reports its own results; (2)~\textbf{counter-signed}, where the delegator independently verifies the result and appends an attestation block; and (3)~\textbf{third-party attested}, where an external verifier (peer verification as defined in our \ldp{} protocol~\citep{prakash2025ldp}, human reviewer, or audit service) signs an attestation block. The \texttt{verification\_status} field maps directly to the verification taxonomy we defined in \ldp{}.

\subsection{Budget Semantics}
\label{sec:budget}

Budget values in \ibct{}s are expressed as integer cents (forced by Biscuit Datalog's lack of floating-point types). Budget fields represent \textbf{per-token authorization ceilings}, not running balances. When Agent~A delegates to Agent~B with \texttt{budget:50}, A asserts that B may spend up to 50 cents on this task. At invocation time, the verifier checks that the declared budget is non-negative; it does not track cumulative spend. Completion blocks record actual cost for audit. Aggregate spend enforcement is the runtime's responsibility, not the token's.

\subsection{Policy Profiles}
\label{sec:policy}

Datalog policies in chained-mode tokens use one of three profiles, ordered by increasing expressiveness and complexity.

\paragraph{Simple profile.} Users specify parameter values (tool allowlist, budget ceiling, maximum depth, expiry) and the library generates canonical Datalog checks. The four canonical templates are normative; implementations MUST generate exactly these patterns to ensure cross-implementation interoperability:

\begin{lstlisting}[style=protocol]
check if tool($t), ["search","browse"].contains($t);
check if budget($b), $b <= 50;
check if depth($d), $d <= 3;
check if time($t), $t <= 2026-03-22T12:00:00Z;
\end{lstlisting}

No Datalog knowledge is required from the user. The Simple profile covers the majority of single-organization delegation scenarios.

\paragraph{Standard profile.} A curated subset of Datalog that permits conjunctive conditions and set membership tests but prohibits recursion and unbounded evaluation. Example policies include trust-domain restrictions (\texttt{check if delegator(\$d), trust\_domain(\$d, \$dom), ["internal"].contains(\$dom)}) and conditional tool access based on delegator attributes.

\paragraph{Advanced profile.} Full Biscuit Datalog with an evaluation depth limit of 1,000 iterations. This profile supports recursive policy rules, cross-block fact derivation, and arbitrary constraint combinations. It is opt-in and intended for enterprise deployments with complex governance requirements. Verifiers that do not support the Advanced profile MUST reject tokens containing non-Standard Datalog rather than silently ignoring policy blocks.

\subsection{Protocol Bindings}
\label{sec:bindings}

\aip{} defines concrete bindings for three transport protocols, ensuring that a single identity and token format works across the agent ecosystem.

\paragraph{\mcp{} binding.} Clients include the \ibct{} in the \texttt{X-AIP-Token} HTTP header on every tool call. The server extracts the token, resolves the issuer's identity document, verifies all signatures, evaluates Datalog policies against the requested tool, and injects the verified identity into the request context. \mcp{} servers declare \texttt{require\_aip: true} in their identity document to reject anonymous calls. Nine structured error codes distinguish authentication failures (HTTP~401: token missing, malformed, expired, signature invalid, identity unresolvable, key revoked) from authorization failures (HTTP~403: scope insufficient, budget exceeded, depth exceeded).

\paragraph{\atoa{} binding.} Agents advertise their \aip{} identity through the \texttt{aip\_identity} field in the \atoa{} agent card~\citep{google2025a2a}. When submitting a task, the caller places the \ibct{} (with a delegation block appended) in the \texttt{aip\_token} field of the task metadata. The receiver verifies the full chain and may further delegate if the remaining depth permits.

\paragraph{HTTP binding.} For generic HTTP APIs, the token is carried in the \texttt{Authorization} header using the \texttt{AIP} scheme: \texttt{Authorization: AIP <token>}. For tokens exceeding 4\,KB, implementations MAY use token-by-reference via the \texttt{X-AIP-Token-Ref} header.

\subsection{Delegation Rules}
\label{sec:delegation}

Six rules govern how \ibct{}s propagate through delegation chains.

\textbf{Scope attenuation only.} Each delegation block MUST be a subset of its parent's capabilities. A block that attempts to widen any scope, increase the budget, or extend the expiry fails cryptographic verification. This is enforced by the Biscuit authorizer, which evaluates all blocks together and rejects any fact that contradicts an earlier attenuation.

\textbf{Bounded depth.} Block~0 declares a \texttt{max\_depth} value (default:~3). Delegation beyond this depth is rejected. The depth counter represents the maximum number of delegation blocks permitted, not the number of hops taken so far.

\textbf{Non-empty context.} Every delegation block MUST include a non-empty \texttt{context} field describing why the delegation is occurring. Verifiers MUST reject tokens with missing or empty context fields. This requirement exists for audit trail integrity: a completed \ibct{} should explain the purpose of each hop.

\textbf{Ephemeral grants for sub-agents.} When an orchestrator spawns a short-lived sub-agent, it generates an Ed25519 keypair, assigns the sub-agent an \texttt{aip:key} identity, and appends a delegation block with narrowed scope and a short TTL (typically minutes). The sub-agent uses the token for its work; the token auto-expires. No DNS registration is needed.

\textbf{Key rotation.} DNS-based identities support zero-downtime rotation through overlapping validity windows in the identity document. Self-certifying identities do not support rotation: the key is the identity, so rotation means creating a new identity.

\textbf{Revocation and mutual authentication.} \aip{} v1 prefers short-lived tokens (under one hour) over revocation infrastructure. For high-security scenarios, identity documents MAY declare a CRL endpoint. Mutual authentication in v1 uses DNS-based TLS verification. Mutual authentication for self-certifying identities is deferred to v2.

\subsection{End-to-End Walkthrough}
\label{sec:walkthrough}

This subsection traces a concrete delegation through the full protocol. A human user asks an orchestrator to research a topic; the orchestrator delegates to a specialist that calls an \mcp{} tool.

\begin{enumerate}[leftmargin=*,nosep]
  \item \textbf{Identity publication.} Agent ``research-analyst'' at \texttt{jamjet.dev} publishes its identity document at \texttt{https://jamjet.dev/.well-known/aip/agents/research-analyst.json}. The document contains the agent's Ed25519 public key, \texttt{max\_depth:~3}, supported protocol bindings (\mcp{} via \texttt{X-AIP-Token}, \atoa{} via \texttt{aip\_identity}), and a self-signature over the RFC~8785 canonical form.

  \item \textbf{Authority token creation.} The human's system creates an authority \ibct{} (Block~0) with: \texttt{issuer:~aip:web:acme.dev/human-system}, scopes \texttt{[tool:search, tool:email]}, \texttt{budget:~500} (cents), \texttt{max\_depth:~3}, and a 30-minute expiry. The block is signed with the human system's Ed25519 private key.

  \item \textbf{Orchestrator delegates to specialist.} The orchestrator appends Block~1: \texttt{delegator:~aip:web:acme.dev/orchestrator}, \texttt{delegatee:~aip:web:jamjet.dev/agents/research-analyst}, scopes narrowed to \texttt{[tool:search]}, budget narrowed to \texttt{100} cents, and context ``research query: climate policy trends.'' The orchestrator signs Block~1 with its own key. The Biscuit authorizer confirms that \texttt{[tool:search]} is a subset of the parent's \texttt{[tool:search, tool:email]} and that 100 $\leq$ 500.

  \item \textbf{Specialist calls \mcp{} tool.} The specialist sends an HTTP request to the search \mcp{} server with the chained token in the \texttt{X-AIP-Token} header. The server extracts the token, resolves \texttt{aip:web:acme.dev/human-system} to fetch the root public key, verifies all Ed25519 signatures (Block~0 through Block~1), evaluates Datalog policies (\texttt{tool:search} is allowed, budget is non-negative, depth $\leq$ 3, token is not expired), and returns search results.

  \item \textbf{Completion block.} The specialist appends Block~2 (completion): \texttt{result\_hash:~sha256:e3b0c4...}, \texttt{verification\_status:~self\_reported}, \texttt{cost\_usd:~0.03}, \texttt{tokens\_used:~1200}. The completed token now records the full chain from human authorization through orchestrator delegation to specialist execution.
\end{enumerate}

At each step, fields checked include: signature validity, scope subset enforcement, budget attenuation, depth bounds, expiry timestamps, and non-empty delegation context. The completed \ibct{} answers who authorized the action, through which agents the delegation flowed, what constraints applied at each hop, and what the outcome was.

\section{Implementation}
\label{sec:impl}

We provide reference implementations of \aip{} in Python (primary SDK) and Rust (reference implementation). Both implement compact and chained modes with full cross-language interoperability. All code is open source under the Apache 2.0 license.\footnote{\url{https://github.com/sunilp/aip}}

The \textbf{Python SDK} (\texttt{aip-sdk}) is organized into three packages. \texttt{aip\_core} provides Ed25519 keypair management (wrapping the \texttt{cryptography} library), \texttt{AipId} parsing for both \texttt{aip:web} and \texttt{aip:key} schemes, and identity document verification including self-signature checks. \texttt{aip\_token} implements \texttt{CompactToken} creation and verification via PyJWT with EdDSA, \texttt{ChainedToken} creation and delegation via \texttt{biscuit-python}, and \texttt{SimplePolicy} Datalog generation from the four canonical templates defined in Section~\ref{sec:policy}. \texttt{aip\_mcp} provides framework-agnostic middleware for HTTP header extraction, token mode detection, and verification.

The \textbf{Rust implementation} is structured as a Cargo workspace with three crates. \texttt{aip-core} handles Ed25519 key operations via \texttt{ed25519-dalek}, identity types, and document verification. \texttt{aip-token} uses manual JWT construction for compact mode and \texttt{biscuit-auth} 6.0 for chained mode, including delegation block appending and Datalog policy evaluation. \texttt{aip-mcp} implements token extraction from HTTP headers, automatic mode detection (JWT header versus Biscuit magic bytes), and scope verification against the requested operation.

Cross-language \textbf{conformance tests} verify that tokens created in Rust can be verified in Python and vice versa, for both compact and chained modes, with identical accept/reject decisions for scope attenuation. Cross-language interoperability means that a token created by the Rust implementation produces identical accept/reject outcomes when verified by the Python implementation, and vice versa, including for malformed tokens and expired claims.

\aip{} serves as the identity layer for the \jamjet{} agent runtime, which also hosts our \ldp{} protocol~\citep{prakash2025ldp} (provenance) and our \dci{} framework~\citep{prakash2025dci} (reasoning). Identity documents link to \ldp{} via the \texttt{extensions} field, and completion blocks link to provenance records via \texttt{ldp\_provenance\_id}.

Table~\ref{tab:impl} summarizes the implementation across both languages.

\begin{table}[t]
\centering
\caption{Implementation summary. LOC counts include library source only (excluding benchmarks and binaries for Rust).}
\label{tab:impl}
\footnotesize
\begin{tabular}{@{}lllr@{}}
\toprule
\textbf{Component} & \textbf{Rust (crate / LOC)} & \textbf{Python (package / LOC)} & \textbf{Tests} \\
\midrule
Identity   & \texttt{aip-core} / 462  & \texttt{aip\_core} / 309  & 6 + 19 \\
Tokens     & \texttt{aip-token} / 601 & \texttt{aip\_token} / 442 & 7 + 27 \\
Middleware & \texttt{aip-mcp} / 151   & \texttt{aip\_mcp} / 76    & 7 + 12 \\
\midrule
\textbf{Total} & \textbf{1,214 LOC} & \textbf{827 LOC} & \textbf{20 + 58} \\
\bottomrule
\end{tabular}
\end{table}

\section{Evaluation}
\label{sec:eval}

This section evaluates \aip{} along four dimensions: compact mode overhead (microbenchmarks), chained mode scaling (microbenchmarks), real-world deployment overhead with LLM inference (real HTTP and Gemini 2.5 Flash), and adversarial security (six attack categories). All benchmarks were run on an Apple M3 Max (macOS 15.3), using the Python and Rust implementations described in Section~\ref{sec:impl}. Compact mode benchmarks used 1,000 iterations; chained mode used 100 iterations per depth level. Real-world experiments used actual HTTP transport and live LLM inference.

\subsection{Hypotheses}
\label{sec:hypotheses}

We state four hypotheses:

\begin{itemize}[leftmargin=*,nosep]
  \item \textbf{H1:} Compact mode adds negligible overhead to \mcp{} tool calls (sub-millisecond over real HTTP).
  \item \textbf{H2:} Chained mode delegation overhead scales linearly with chain depth, both in token size and verification latency.
  \item \textbf{H3:} \aip{} overhead is negligible relative to real LLM inference in multi-agent delegation.
  \item \textbf{H4:} \aip{} detects attacks that unsigned and plain JWT deployments miss.
\end{itemize}

\subsection{Microbenchmarks (H1, H2)}
\label{sec:microbench}

\paragraph{Compact mode (H1).} Table~\ref{tab:compact} reports create and verify latencies for compact \ibct{}s. Rust verification averages 0.049\,ms (p99: 0.059\,ms); Python verification averages 0.189\,ms (p99: 0.231\,ms). Both are well below 1\,ms. Token size is 356 bytes in both implementations, confirming identical wire format. Rust is approximately 3.9$\times$ faster for verification, which reflects the overhead of Python's interpreter and FFI boundary to the \texttt{cryptography} library.

\begin{table}[t]
\centering
\caption{Compact mode benchmarks (1,000 iterations). Mean and p99 latencies in milliseconds.}
\label{tab:compact}
\footnotesize
\begin{tabular}{@{}lcccc@{}}
\toprule
\textbf{Operation} & \textbf{Rust mean} & \textbf{Rust p99} & \textbf{Python mean} & \textbf{Python p99} \\
\midrule
Create   & 0.018 & 0.022 & 0.086 & 0.110 \\
Verify   & 0.049 & 0.059 & 0.189 & 0.231 \\
\midrule
\multicolumn{5}{@{}l@{}}{\textit{Token size: 356 bytes (identical across languages)}} \\
\bottomrule
\end{tabular}
\end{table}

\paragraph{Chained mode scaling (H2).} Table~\ref{tab:chained} reports token size and verification latency at depths 0--5. Token size grows linearly (340--380 bytes per block depending on the library's Biscuit serialization). Verification latency also grows linearly. Rust is slower than Python for chained verification because \texttt{biscuit-auth} 6.0 performs more thorough Datalog evaluation than the \texttt{biscuit-python} binding. Both remain under 1\,ms at the maximum recommended depth.

\begin{table}[t]
\centering
\caption{Chained mode scaling (100 iterations per depth). Token size in bytes; verify latency in milliseconds.}
\label{tab:chained}
\footnotesize
\begin{tabular}{@{}rrrcc@{}}
\toprule
\textbf{Depth} & \textbf{Size (Py)} & \textbf{Size (Rust)} & \textbf{Python verify} & \textbf{Rust verify} \\
\midrule
0 &  444 &  520 & 0.110 & 0.188 \\
1 &  832 &  940 & 0.178 & 0.292 \\
2 & 1,172 & 1,316 & 0.248 & 0.403 \\
3 & 1,512 & 1,696 & 0.315 & 0.516 \\
4 & 1,856 & 2,072 & 0.379 & 0.625 \\
5 & 2,196 & 2,448 & 0.447 & 0.745 \\
\bottomrule
\end{tabular}
\end{table}

\paragraph{Security conformance.} Scope attenuation was correctly enforced in 100/100 test cases. Each test attempted to widen scope beyond the parent block's authority (adding tools not in the parent allowlist, increasing budget, extending expiry). All 100 attempts were rejected by both the Python and Rust verifiers.

\paragraph{Discussion.} Across both modes, all verification latencies remain sub-millisecond. A depth-5 chained token (the recommended maximum) is under 2.5\,KB, well within the 8\,KB HTTP header limit. \aip{} verification is not a bottleneck for agent tool calls.

\subsection{Real-World MCP Deployment (H1, H3)}
\label{sec:real-mcp}

To validate compact mode overhead under realistic conditions, we measured \aip{} token creation and verification over real HTTP (localhost) with 100 iterations per condition. Table~\ref{tab:real-mcp} reports mean and p99 latencies for three conditions: no authentication, \aip{} compact mode, and \aip{} chained mode.

\begin{table}[t]
\centering
\caption{Real MCP tool call latency over HTTP (100 iterations). All values in milliseconds.}
\label{tab:real-mcp}
\footnotesize
\begin{tabular}{@{}lcccc@{}}
\toprule
\textbf{Condition} & \textbf{Mean} & \textbf{p50} & \textbf{p99} & \textbf{Overhead vs.\ No Auth} \\
\midrule
No authentication  & 0.301 & 0.294 & 0.404 & -- \\
AIP compact        & 0.523 & 0.507 & 0.819 & +0.222\,ms \\
AIP chained        & 0.481 & 0.476 & 0.819 & +0.180\,ms \\
\bottomrule
\end{tabular}
\end{table}

Both \aip{} modes add sub-millisecond overhead to real HTTP tool calls. Compact mode adds 0.222\,ms (73.9\% relative increase over a 0.301\,ms baseline), while chained mode adds 0.180\,ms (59.8\%). The absolute overhead in both cases is under a quarter of a millisecond. Chained mode is slightly faster than compact because the Biscuit library amortizes key operations across blocks, while compact mode pays the full JWT signature cost per call.

\subsection{Multi-Agent Delegation with LLM Inference (H3)}
\label{sec:real-llm}

The critical question for production deployment is whether \aip{} overhead matters relative to LLM inference. We measured a complete orchestrator-to-specialist delegation with real Gemini 2.5 Flash inference over 10 iterations. Table~\ref{tab:real-llm} breaks down the time budget.

\begin{table}[t]
\centering
\caption{Multi-agent delegation with Gemini 2.5 Flash (10 iterations). All values in milliseconds.}
\label{tab:real-llm}
\footnotesize
\begin{tabular}{@{}lrrr@{}}
\toprule
\textbf{Component} & \textbf{Mean} & \textbf{Min} & \textbf{Max} \\
\midrule
AIP create         &    0.385 &   0.146 &    0.934 \\
AIP delegate       &    0.510 &   0.248 &    0.820 \\
AIP verify         &    1.455 &   0.650 &    2.340 \\
\midrule
\textit{AIP total} &   \textit{2.351} &   &    \\
\midrule
LLM orchestrator   & 1,019.886 & 832.813 & 1,293.494 \\
LLM specialist     & 1,727.052 & 1,557.999 & 1,897.933 \\
\midrule
\textit{LLM total} & \textit{2,747.038} &  &  \\
\midrule
\textbf{End-to-end total} & \textbf{2,749.476} & \textbf{2,500.0} & \textbf{3,010.7} \\
\midrule
\textbf{AIP as \% of total} & \textbf{0.086\%} & & \textbf{p99: 0.127\%} \\
\bottomrule
\end{tabular}
\end{table}

\aip{} protocol overhead is less than 0.13\% of total end-to-end latency even at p99. The mean overhead is 2.351\,ms against a mean total of 2,749\,ms. LLM inference dominates by three orders of magnitude: the orchestrator call averages 1,020\,ms and the specialist averages 1,727\,ms, together accounting for 99.91\% of total time. The structural advantage of \aip{} is clear: because verification is purely local (no authorization server round-trip), protocol overhead scales with cryptographic operations, not network latency.

\subsection{Adversarial Security Evaluation (H4)}
\label{sec:adversarial}

We evaluated \aip{} against six attack categories, with 100 iterations per attack (600 total attempts), under three deployment configurations: \aip{} (compact + chained), unsigned (no authentication), and plain JWT (standard signed JWT without \aip{} delegation semantics). Table~\ref{tab:adversarial} reports rejection rates.

\begin{table*}[t]
\centering
\caption{Adversarial security evaluation (100 iterations per attack, 600 total). Rejection rate shown as fraction of attempts blocked.}
\label{tab:adversarial}
\footnotesize
\begin{tabular}{@{}p{3.2cm}p{5.5cm}cccp{3.8cm}@{}}
\toprule
\textbf{Attack} & \textbf{Description} & \textbf{AIP} & \textbf{Unsigned} & \textbf{JWT} & \textbf{AIP Unique Defense} \\
\midrule
Scope widening & Agent invokes \texttt{tool:email} when only \texttt{tool:search} was delegated & 100\% & 0\% & 100\% & Chained Datalog checks enforce scope at every delegation hop \\
\addlinespace
Depth violation & Second delegation attempted when \texttt{max\_depth=1} & 100\% & 0\% & 0\% & \texttt{max\_depth} in authority block prevents unbounded delegation \\
\addlinespace
Expired token replay & Token with \texttt{exp=now-60s} presented for verification & 100\% & 0\% & 100\% & Both compact (JWT \texttt{exp}) and chained (Datalog time check) enforce expiry \\
\addlinespace
Wrong key verification & Token signed by key A, verified against unrelated key B & 100\% & 0\% & 100\% & Ed25519 signature binding at every layer \\
\addlinespace
Empty context (audit evasion) & Delegation with empty context string to evade audit trail & 100\% & 0\% & 0\% & Mandatory non-empty context on every delegation enforces audit provenance \\
\addlinespace
Token forgery & Base64 token tampered (character flip), then verified & 100\% & 0\% & 100\% & Biscuit signature covers every block; any tampering is detected \\
\midrule
\textbf{Total (600 attempts)} & & \textbf{100\%} & \textbf{0\%} & \textbf{67\%} & \\
\bottomrule
\end{tabular}
\end{table*}

\aip{} rejected 600/600 attack attempts across all six categories. Unsigned deployments rejected 0/600, confirming that without any authentication layer, all attacks succeed. Plain JWT deployments caught four of six attack types (scope widening, expired token replay, wrong key verification, and token forgery) but missed two: depth violation and empty context audit evasion. These two failures are structurally important. Standard JWTs have no concept of delegation depth, so there is no mechanism to enforce \texttt{max\_depth} constraints. JWTs also have no mandatory context requirement, so an agent can delegate silently without recording an audit trail. \aip{} uniquely defends against depth violation and audit evasion through mandatory delegation context and bounded depth in the authority block, attacks that neither unsigned nor JWT-only deployments can detect.

\subsection{Evidence Summary}
\label{sec:evidence}

\textbf{H1 supported:} Compact verification is 0.049\,ms (Rust) and 0.189\,ms (Python) in microbenchmarks. In a real MCP deployment over HTTP, compact mode adds 0.222\,ms and chained mode adds 0.180\,ms, both sub-millisecond. Token size is 356 bytes (compact) and scales linearly in chained mode.

\textbf{H2 supported:} Microbenchmarks confirm linear scaling. Token size grows at 340--380 bytes per block. Verification latency grows linearly, reaching 0.745\,ms (Rust) at depth~5. Scope attenuation was enforced in 100/100 test cases.

\textbf{H3 supported:} With real Gemini 2.5 Flash inference, \aip{} adds 2.351\,ms mean overhead against 2,749\,ms total latency, representing 0.086\% of end-to-end time (0.127\% at p99). Protocol overhead is three orders of magnitude smaller than LLM inference.

\textbf{H4 supported:} \aip{} rejected 600/600 adversarial attempts across six attack categories. Plain JWT missed two categories (depth violation, empty context audit evasion) that \aip{}'s chained delegation model uniquely detects. Unsigned deployments missed all six.

\section{Threat Model}
\label{sec:threat}

This section defines seven adversary classes relevant to agent delegation and characterizes what \aip{} prevents, detects, or does not address for each.

\paragraph{Malicious delegatee (scope widening).} A delegatee that attempts to widen its granted scope beyond the parent block's authority is prevented by Biscuit's check-all-blocks semantics: the authorizer evaluates every block in the chain and rejects any fact that contradicts an earlier attenuation. This is a cryptographic guarantee, not a runtime policy. However, a delegatee acting maliciously within its granted scope (e.g., calling permitted tools for harmful purposes) is out of scope for the token layer.

\paragraph{Compromised identity host (DNS hijack).} An attacker who compromises a DNS-based identity host can serve a forged identity document. \aip{} does not prevent this attack, but the document self-signature (Section~\ref{sec:identity}) enables detection: a forged document will fail signature verification unless the attacker also possesses the agent's private key. If the attacker holds the actual private key, the identity is fully compromised and \aip{} provides no defense.

\paragraph{Replay attacker.} An attacker who captures and replays an \ibct{} is partially mitigated by short-lived token expiry. Tokens with expiry timestamps that have passed are rejected during verification. However, real-time replay within the token's TTL window is not prevented by the token alone; transport-layer protections (TLS, nonce binding) provide the primary defense.

\paragraph{Dishonest verifier.} A verifier that skips signature checks, ignores policy evaluation, or accepts expired tokens can grant unauthorized access. \aip{} assumes the verifier is trusted and correctly implemented. Verifier compliance is an operational concern addressed through conformance testing and reference implementations, not a property the protocol can enforce cryptographically.

\paragraph{Fraudulent completion block.} Completion blocks are self-reported by default (Section~\ref{sec:completion}): the executing agent signs its own result claim. \aip{} provides attribution binding (the signature identifies who made the claim) but does not prevent a dishonest agent from misrepresenting results. Counter-signing and third-party attestation provide stronger guarantees but are optional in v1.

\paragraph{Colluding sub-agents.} Two or more agents in a delegation chain that collude to misuse granted authority are not prevented by \aip{}. The delegation chain records who delegated to whom, providing partial detection through audit trail analysis, but collusion within granted scopes is outside the protocol's threat model.

\paragraph{DoS via policy complexity.} An attacker who crafts a token with deeply nested Datalog rules to exhaust verifier CPU is partially mitigated by the policy profile system. The Simple profile generates only four canonical check templates with bounded evaluation. The Advanced profile permits full Datalog but caps evaluation at 1,000 iterations. Verifiers that enforce profile restrictions bound their worst-case evaluation cost.

\paragraph{Summary.} Table~\ref{tab:threat-summary} provides a compact overview of the threat model.

\begin{table}[t]
\centering
\caption{Threat model summary. Each row maps an adversary to what \aip{} prevents, detects, or does not address.}
\label{tab:threat-summary}
\footnotesize
\setlength{\tabcolsep}{3pt}
\begin{tabular}{@{}p{2.4cm}p{1.8cm}p{1.8cm}p{1.8cm}@{}}
\toprule
\textbf{Adversary} & \textbf{Prevented} & \textbf{Detected} & \textbf{Not Addressed} \\
\midrule
Scope widening      & Datalog check-all-blocks & --          & Misuse within scope \\
DNS hijack          & --          & Self-signature mismatch & Key compromise \\
Replay              & Expiry check & --          & Real-time replay in TTL \\
Dishonest verifier  & --          & Conformance tests & Skipped checks \\
Fraudulent completion & --        & Attribution binding & Fabricated results \\
Colluding agents    & --          & Audit trail & In-scope collusion \\
DoS via Datalog     & Profile limits & Iteration cap & Novel eval exploits \\
\bottomrule
\end{tabular}
\end{table}

\section{Limitations}
\label{sec:limitations}

\textbf{Limited production deployment data.} While Section~\ref{sec:eval} includes real HTTP transport and real LLM inference (Gemini 2.5 Flash), the evaluation was conducted on a single machine with localhost networking. We have not deployed \aip{} in a production multi-agent system with real user traffic, cross-datacenter networking, or sustained load. The performance characteristics are representative of cryptographic and protocol overhead but do not capture production networking conditions.

\textbf{Datalog verifier complexity.} The Biscuit Datalog verifier in chained mode is a potential attack surface. A maliciously crafted token with deeply nested or recursive policy rules could consume excessive CPU during verification. The Advanced policy profile caps evaluation at 1,000 iterations, and the Simple profile (which generates only the four canonical templates) eliminates this risk entirely. We recommend Simple as the default for all deployments.

\textbf{Completion blocks are self-reported.} By default, completion blocks are signed by the executing agent, which is the party whose work is being reported (Section~\ref{sec:completion}). In our earlier work on the Provenance Paradox~\citep{prakash2025provenance}, we demonstrated that self-claimed quality systematically selects the worst delegates, reinforcing the need for independent verification. Counter-signing and third-party attestation exist as trust escalation options but are not enforced in v1. A dishonest agent can misrepresent its result hash or cost without detection by the cryptographic layer alone.

\textbf{No revocation infrastructure.} \aip{} v1 relies on short-lived tokens (under one hour) rather than revocation checking. For scenarios requiring immediate key compromise response, identity documents may declare a CRL endpoint, but no reference implementation enforces CRL checks. Full revocation infrastructure is deferred to v2.

\textbf{DNS trust anchor vulnerabilities.} DNS-based identities (\texttt{aip:web}) inherit DNS vulnerabilities including hijacking and cache poisoning. The document self-signature (Section~\ref{sec:identity}) detects content tampering even when the transport is compromised, but this defense requires verifiers to actually check the signature. Verifier compliance is an operational concern, not a cryptographic guarantee.

\textbf{No post-quantum readiness.} \aip{} v1 uses Ed25519 exclusively. No post-quantum signature scheme is supported. Algorithm agility (supporting multiple signature algorithms with negotiation) is deferred to a future version. The choice of Ed25519-only simplifies implementation and avoids algorithm negotiation complexity, but creates a migration burden if quantum-capable adversaries become a practical threat.

\section{Conclusion}
\label{sec:conclusion}

We introduced Invocation-Bound Capability Tokens (\ibct{}s), implemented in the Agent Identity Protocol (\aip{}). \ibct{}s fuse identity, attenuated authorization, and provenance into a single append-only chain operating in two wire formats: compact mode (JWT, 356 bytes, 0.049\,ms verification in Rust) for single-hop \mcp{} calls, and chained mode (Biscuit with Datalog policies, linear scaling at 340--380 bytes per block) for multi-hop delegation with audit trails. Reference implementations in Python and Rust demonstrate cross-language interoperability across 78 tests.

In a real MCP deployment over HTTP, \aip{} adds 0.22\,ms overhead over unauthenticated baselines. In a real multi-agent deployment with Gemini 2.5 Flash inference, \aip{} adds 2.35\,ms of overhead (0.086\% of total latency), confirming that identity verification is not a bottleneck. Adversarial evaluation across 600 attack attempts shows 100\% rejection, with two categories (delegation depth violation and audit evasion) uniquely caught by \aip{}'s chained delegation model.

Three next steps follow. First, the \atoa{} protocol binding needs validation against production cross-organization delegation workflows. Second, the provenance bridge between \aip{} completion blocks and \ldp{}~\citep{prakash2025ldp} should be formalized into a joint specification. Third, a controlled comparison against a real OAuth 2.1 baseline (rather than no-auth and plain JWT) would quantify the latency and expressiveness tradeoffs in production settings.


\begin{thebibliography}{22}
\providecommand{\natexlab}[1]{#1}
\providecommand{\url}[1]{\texttt{#1}}
\expandafter\ifx\csname urlstyle\endcsname\relax
  \providecommand{\doi}[1]{doi: #1}\else
  \providecommand{\doi}{doi: \begingroup \urlstyle{rm}\Url}\fi

\bibitem[{Anthropic}(2024)]{anthropic2024mcp}
{Anthropic}.
\newblock Model context protocol specification.
\newblock \url{https://modelcontextprotocol.io/}, 2024.

\bibitem[Birgisson et~al.(2014)Birgisson, Politz, Erlingsson, Taly, Vrable, and
  Lentczner]{birgisson2014macaroons}
Arnar Birgisson, Joe~Gibbs Politz, \'Ulfar Erlingsson, Ankur Taly, Michael
  Vrable, and Mark Lentczner.
\newblock Macaroons: Cookies with contextual caveats for decentralized
  authorization in the cloud.
\newblock In \emph{Network and Distributed System Security Symposium}, 2014.

\bibitem[Cruz(2026)]{ietf2026aap}
A.~Cruz.
\newblock Agent authorization profile ({AAP}) for {OAuth} 2.0.
\newblock IETF Internet-Draft draft-aap-oauth-profile-01, 2026.

\bibitem[{Eclipse Foundation}(2024)]{biscuit2024spec}
{Eclipse Foundation}.
\newblock Biscuit authorization token specification.
\newblock \url{https://www.biscuitsec.org/}, 2024.

\bibitem[{Google}(2025)]{google2025a2a}
{Google}.
\newblock Agent-to-agent ({A2A}) protocol.
\newblock \url{https://a2a-protocol.org/}, 2025.

\bibitem[Goswami(2026)]{ietf2026agenticjwt}
Abhishek Goswami.
\newblock Secure intent protocol: {JWT} compatible agentic identity and
  workflow management.
\newblock IETF Internet-Draft draft-goswami-agentic-jwt-00, 2026.

\bibitem[Hardt et~al.(2026)Hardt, Parecki, and Lodderstedt]{ietf2024oauth21}
Dick Hardt, Aaron Parecki, and Torsten Lodderstedt.
\newblock The {OAuth} 2.1 authorization framework.
\newblock IETF Internet-Draft draft-ietf-oauth-v2-1-15, 2026.

\bibitem[{IETF}(2026)]{ietf2026scim}
{IETF}.
\newblock {SCIM} profile for {AI} agents.
\newblock IETF Internet-Draft, 2026.
\newblock Agent provisioning lifecycle management.

\bibitem[{IETF OAuth Working Group}(2026)]{ietf2026txntokens}
{IETF OAuth Working Group}.
\newblock Transaction tokens for agents.
\newblock IETF Internet-Draft draft-oauth-transaction-tokens-for-agents-04,
  2026.

\bibitem[Kasselman et~al.(2026)Kasselman, Lombardo, Rosomakho, and
  Campbell]{ietf2026aims}
Pieter Kasselman, John Lombardo, Yaroslav Rosomakho, and Brian Campbell.
\newblock {AI} agent authentication and authorization.
\newblock IETF Internet-Draft draft-klrc-aiagent-auth-00, 2026.

\bibitem[{Knostic}(2025)]{knostic2025scan}
{Knostic}.
\newblock {MCP} server security scan, 2025.
\newblock Scan of approximately 2,000 MCP servers found all lacked
  authentication.

\bibitem[{Mastercard}(2025)]{mastercard2025intent}
{Mastercard}.
\newblock Verifiable intent for agent commerce, 2025.
\newblock Cryptographic audit trail for agent-initiated transactions.

\bibitem[N{\'e}methi(2026)]{ietf2026aid}
Bal{\'a}zs N{\'e}methi.
\newblock Agent identity and discovery ({AID}).
\newblock IETF Internet-Draft draft-nemethi-aid-agent-identity-discovery-00,
  2026.

\bibitem[Ni and Liu(2026)]{ietf2026wimse}
Y.~Ni and C.~P. Liu.
\newblock {WIMSE} applicability for {AI} agents.
\newblock IETF Internet-Draft draft-ni-wimse-ai-agent-identity-02, 2026.

\bibitem[Prakash(2025{\natexlab{a}})]{prakash2025dci}
Sunil Prakash.
\newblock Deliberative collective intelligence: A framework for structured
  multi-agent reasoning.
\newblock \emph{arXiv preprint arXiv:2603.11781}, 2025{\natexlab{a}}.

\bibitem[Prakash(2025{\natexlab{b}})]{prakash2025ldp}
Sunil Prakash.
\newblock Identity-aware delegation for {LLM} agents: A protocol-level approach
  with structured provenance.
\newblock \emph{arXiv preprint arXiv:2603.08852}, 2025{\natexlab{b}}.

\bibitem[Prakash(2025{\natexlab{c}})]{prakash2025provenance}
Sunil Prakash.
\newblock The provenance paradox: Why self-claimed quality systematically
  selects the worst delegates.
\newblock \emph{arXiv preprint arXiv:2603.18043}, 2025{\natexlab{c}}.

\bibitem[Rundgren et~al.(2020)Rundgren, Jordan, and Erdtman]{rfc8785}
Anders Rundgren, Bret Jordan, and Samuel Erdtman.
\newblock {JSON} canonicalization scheme ({JCS}).
\newblock RFC 8785, \url{https://www.rfc-editor.org/rfc/rfc8785}, 2020.

\bibitem[{SPIFFE Project}(2024)]{spiffe2024}
{SPIFFE Project}.
\newblock Secure production identity framework for everyone ({SPIFFE}).
\newblock \url{https://spiffe.io/}, 2024.

\bibitem[Toma{\v{s}}ev et~al.(2026)Toma{\v{s}}ev, Franklin, and
  Osindero]{deepmind2026delegation}
Nenad Toma{\v{s}}ev, Matija Franklin, and Simon Osindero.
\newblock Intelligent {AI} delegation.
\newblock \emph{arXiv preprint arXiv:2602.11865}, 2026.

\bibitem[{UCAN Working Group}(2024)]{ucan2024spec}
{UCAN Working Group}.
\newblock User controlled authorization networks ({UCAN}) specification.
\newblock \url{https://ucan.xyz/specification/}, 2024.

\bibitem[{World Wide Web Consortium}(2026)]{w3c2026did}
{World Wide Web Consortium}.
\newblock Decentralized identifiers ({DIDs}) v1.1.
\newblock \url{https://www.w3.org/TR/did-1.1/}, 2026.
\newblock W3C Candidate Recommendation.

\end{thebibliography}
\end{document}